\documentclass{aa}
\usepackage{graphics}
\begin{document}
\thesaurus{02.01.1; 02.18.5; 08.19.5 Cas~A; 13.07.3; 13.18.3}
\title{ On the gamma-ray fluxes expected from Cassiopeia~A}
\author{ A. M. Atoyan \inst{1,2}
\and F. A. Aharonian \inst1
\and R. J. Tuffs \inst1
\and H. J. V\"olk \inst1}
\offprints{Felix.Aharonian@mpi-hd.mpg.de}
\institute{Max Planck Institut f\"ur Kernphysik, Saupfercheckweg 1,
           D--69117 Heidelberg, Germany
	   \and Yerevan Physics Institute, 375036 Yerevan, Armenia}	   
\date{Received 6 July ; accepted 12 Dec 1999}
%\titlerunning{Gamma-rays from Cas A}
\maketitle

\begin{abstract}
Based on the results of our previous study 
of the broad band synchrotron emission of Cas~A in the framework of a spatially
inhomogeneous model,  
we calculate the fluxes of $\gamma$-radiation that can be expected from this
supernova remnant in different energy bands. We show that at energies 
below 10 GeV  $\gamma$-ray fluxes detectable by forthcoming space-borne
detectors should be inevitably produced due to bremsstrahlung of radio emitting
electrons. We predict that the power-law index of the photon flux in the GeV 
region should be hard, close to the index $\beta_{\rm acc}\sim 2.2$ expected
for the acceleration spectrum of electrons in compact bright radio structures. 
Photon fluxes accessible to future
ground-based $\gamma$-ray detectors could also be expected at
very high energies. The fluxes to be expected due to bremsstrahlung and 
inverse Compton radiation of relativistic electrons at TeV energies should 
be very steep, and strongly dependent on the characteristic magnetic 
fields in Cas~A. We could expect also significant fluxes of $\pi^0$-decay 
$\gamma$-rays produced by relativistic protons which presumably are also 
accelerated in Cas~A. 
These fluxes may extend with a hard spectrum beyond TeV energies as far as
the protons could be accelerated to energies $\geq 100\,\rm TeV$. The hardness
of the $\gamma$-ray spectrum at TeV energies could in principle allow one to 
distinguish between electronic and hadronic origins of those $\gamma$-rays.
We discuss also other implications, such as relativistic particle content,
or physical parameters in the source, that could be derived from future 
$\gamma$-ray flux measurements in different energy bands.

\end{abstract}

\section{Introduction}

The shell type supernova remnant (SNR) Cassiopeia A is a prominent 
source of nonthermal radiation in the Galaxy. It is the brightest and 
one of the best 
studied radio sources (e.g. Bell 1975; Tuffs 1986; Braun et al. 1987; 
Anderson et al. 1991; Kassim et al. 1995; etc.), with synchrotron 
radiation fluxes observed also at sub-millimeter wavelengths (Mezger et al. 
1986), and probably even further in the near infrared (Tuffs et al. 1997)
and at hard X-ray energies (Allen et al. 1997; Favata et al. 1997). The
powerful 
radio emission of Cas~A implies a total energy in relativistic electrons
of order $10^{48}\,\rm erg$ or higher (e.g. Chevalier et al. 1978; 
Anderson et al. 1991). A synchrotron origin of hard X-rays implies 
that relativistic electrons are accelerated up to energies 
of a few tens of TeV (Allen et al. 1997; Favata et al. 1997).
Thus, one should also expect the production of nonthermal $\gamma$-rays 
in Cas~A, due to the bremsstrahlung and inverse Compton (IC) mechanisms, 
extending possibly beyond TeV energies. However, in the MeV $\gamma$-ray 
domain only the $^{44}\rm Ti$ line emission at an energy of 
$1.157\,\rm MeV$ has been detected by COMPTEL (Iyudin et al. 1994). 
High energy $\gamma$-ray observations of Cas~A by EGRET resulted in the 
flux upper limit $I(>100\,\rm MeV)\leq 1.2\times 10^{-7}\,\rm ph/cm^2 s$ 
(Esposito et al. 1996). At TeV energies flux upper limits have been given 
by the Whipple (Lessard et al. 1999) and CAT (Goret et al. 1999)
collaborations. A tentative detection of a weak signal at a 
significance level of $\simeq 5\sigma$ has been recently reported by 
the HEGRA collaboration (P\"uhlhofer et al. 1999). 

Gamma-ray fluxes expected from Cas~A have been earlier 
calculated by Cowsik \& Sarkar (1980) who have derived a lower
limit to the mean magnetic field in the shell of Cas~A, 
$B_0\geq 8\times 10^{-5}\,\rm G$, comparing the 
 bremsstrahlung flux of radio emitting electrons with the upper flux limit 
$I(>100\,\rm MeV)\leq 1.1\times 10^{-6}\,\rm ph/cm^2 s$ of 
SAS-2 (Fichtel et al. 1975) and COS~B detectors. Calculations have been 
done in a commonly used `single-zone' model approximation, which assumes 
a spatially homogeneous 
source containing magnetic field, relativistic electrons and gas.
In the same single-zone approximation, broad-band fluxes 
of $\gamma$-rays expected from Cas~A have been
recently calculated by Ellison et al. (1999) and Goret et al. (1999).

In our previous work (Atoyan et al. 1999, hereafter Paper 1) we have carried 
out detailed modelling of the synchrotron radiation of Cas~A from the radio to 
the X-ray bands in order to understand the relativistic electron content 
in the source. The principal feature of our study in Paper 1 was that we 
have proposed a {\it spatially inhomogeneous} model, consisting of two zones,
that distinguishes between compact, bright steep-spectrum 
radio knots and the bright fragmented  
radio ring on the one hand, and the remainder of the shell - the 
diffuse `plateau' - on the other hand. 
In this paper, using the model parameters derived from the interpretation
of synchrotron fluxes observed, we calculate the fluxes of $\gamma$-rays 
predicted in the framework of the spatially inhomogeneous study of this
SNR, and discuss implications which can be derived from   
future $\gamma$-ray observations of this SNR in different energy bands.
In section 2, after a brief overview of the basic features of the model, we
study the fluxes to be expected for  
$\gamma$-ray energies up to $E\sim 10-100\,\rm GeV$. This emission is
contributed mostly by bremsstrahlung of those electrons which are responsible 
for the synchrotron emission in the radio-to-infrared bands. 
In section 3 we discuss the fluxes expected in the very high energy
region, $E\geq 100\,\rm GeV$, where both bremsstrahlung and inverse Compton
radiation of relativistic electrons are important. In addition, a significant
contribution to the total fluxes could be due to $\pi^0$-decay 
$\gamma $-rays. These should originate from 
relativistic protons which presumably are also accelerated
in Cas~A. A summary and the conclusions are contained in section 4.

\section{High energy bremsstrahlung of radio electrons}

\subsection{Overview of the model}

In Paper 1 we have shown that the basic features of the observed 
radio emission, in particular,

\noindent
(a) spectral turnover of the energy fluxes 
$J_{\nu}\sim \nu^{-\alpha}$ at frequencies below 20\,MHz, 

\noindent
(b) mean spectral index $\alpha\approx 0.77$ of the total flux 
at higher frequencies up to 
$\nu\leq 30\,\rm GHz$ (e.g. Baars et al. 1977) which then flattens to 
$\alpha\approx 0.65$ in the submillimeter domain (Mezger et al. 1986), 

\noindent
(c) the spread of spectral indices of the radio knots from 
$\alpha\sim 0.6$ to $\alpha\sim 0.9$ (Anderson and Rudnick, 1996),

\vspace{1mm}

\noindent
and other characteristics can be explained in the framework of a model 
that takes into account the effects of energy dependent propagation of 
relativistic electrons in a spatially nonuniform remnant. We have considered 
the simplest inhomogeneous model, consisting of two zones
with essentially different spatial densities of relativistic
electrons. The model allows a rather clear qualitative and quantitative
distinction between compact bright radio structures in Cas~A like 
the fragmented radio ring and the radio knots on the 
one hand, and the remainder of the radio emitting shell -- 
the diffuse radio `plateau' region enclosed between the radio ring at 
$R_{\rm ring}\approx 1.7\,\rm pc$ and the blast wave at $R_0\approx 2.5\,\rm
pc$ (for a distance  $d=3.4\,\rm kpc$ to Cas~A, cf. Reed et al. 1995) -- 
on the other. The model predicts that the 
spatial density of relativistic electrons in the compact radio structures,
 which all together constitute what we termed zone 1, is much higher 
than in the surrounding plateau 
region, termed zone 2.  Thus, zone 1 with a total volume $V_1$ much
smaller that the volume $V_0$ of the shell is actually merged into 
 zone 2 with a volume $V_2=V_0-V_1 \approx V_0$.

In Paper 1 we have derived, by spatial integration of the diffusion equation
for relativistic electrons,
 the set of Leaky-box type equations for 
the overall energy distribution functions of the electrons 
$N_1(E,t)$ and $N_2(E,t)$ in zones 1 and 2, respectively. We do not consider
the acceleration process itself, but assume that the accelerated electrons
are injected into zone 1 with some rate $Q(E,t)$, and study the
effects of the energy-dependent diffusive propagation of the particles on their
spectra $N_{1,2}(E,t)$. Such a phenomenological approach is  
justified if particle acceleration occurs in a volume significantly smaller 
than the volume where the bulk of the emission of zone 1 is produced.  
The equation for zone 1 then reads:
\begin{equation}
\frac{\partial N_1}{\partial t} = \frac{\partial (\,P_1\,N_1\,)}
{\partial E} \, - \, \frac{N_1}{\tau_{\rm esc}}\,  
+\, \frac{V_1\,N_2}{V_2\,\tau_{\rm dif}}\, +\, Q \; ,
\end{equation}
where $P_{1}$ represents the energy loss rate of the electrons in
zone 1. The second term on the right hand side of Eq.(1)
describes the escape of the 
electrons into the plateau region on timescales
\begin{equation}
\tau_{\rm esc}(E) = \left[ \frac{1}{\tau_{\rm dif}(E)}
 +\frac{1}{\tau_{\rm c}}\right]^{-1}\; .
\end{equation}
The term $\tau_{\rm c}\sim 2R_1/u_1$, where $R_1\sim (0.03-0.1)\,\rm pc$ 
is a typical radius of the compact radio components and 
$u_1$ is the fluid speed in zone 1, corresponds to the convective escape 
time which does not depend on the electron energy. The principal energy
dependent term in Eq.(2) is the diffusive escape time: 
\begin{equation}
\tau_{\rm dif}(E) = \tau_{\ast} \,(E/E_{\ast})^{- \delta}
\, +\, \tau_{\rm min}\; ,
\end{equation}
where normalization to a typical energy $E_\ast =1\,\rm GeV$ of radio 
emitting electrons is used, and $\tau_{\rm min}=2R_1/c$ takes into account
that the escape time cannot be shorter than the light travel time. 

The third term on the right hand side of Eq.(1) takes into account that  
 the diffusive propagation and escape of particles is effectively
possible only in the directions opposite to the gradients of their spatial
density, i.e. from zone 1 with a high concentration
of radio electrons into zone 2.      
In a standard power-law approximation
for the above source function 
\begin{equation}
Q(E) = \, Q_0\, E^{-\beta_{\rm acc} }\exp(-E/E_{\rm c})\; ,
\end{equation}
the spectrum of the accelerated electrons can be hard, 
$\beta_{\rm acc} \simeq 2.2$. 
 However the energy distribution $N_{1}(E)$ of radio
electrons in zone 1 becomes steeper than  $Q(E)$ on the timescale 
$\tau_{\rm esc}(E)$ of electron escape from zone 1 into zone 2. It is 
important that, as shown in Paper 1, the degree of steepening at 
energy $E$ essentially depends on 
the ratio (or `the gradient') of the energy densities of the electrons
$N_1(E)/V_1$ to $N_2(E)/V_2$ in these two zones. The 
maximum steepening, corresponding to the spectrum 
$N_{1}(E) \simeq Q(E)\times\tau_{\rm esc}(E) \propto
E^{-\beta_{\rm max}}$ with $\beta_{\rm max}=\beta_{\rm acc}+\delta$, is 
reached only if this gradient is very high, and there is {\it no steepening
at all} if the energy densities become equal. Therefore, although the 
two-zone model is also rather simplified, as compared with the reality,
 because it assumes the  
same electron density for all components in zone 1, it allows a qualitative
explanation for the variations of the power-law indices of individual radio 
knots from $\alpha_{\rm min} \simeq 0.6$ to  
$\alpha_{\rm max} \simeq 0.9-0.95$ (Anderson \& Rudnick 1996; see also
 Wright et al. 1999) assuming
an efficient electron acceleration with the same hard power-law spectrum
$\beta_{\rm acc}=1+2\,\alpha_{\rm min}\simeq 2.2$ and a diffusive escape 
 time with a parameter $\delta \simeq 0.6-0.7$. These variations
do not then necessarily
imply that particle acceleration significantly varies across the remnant
(Wright et al. 1999), but can be explained rather by differences in 
the local gradients, with respect to the surrounding plateau, 
of the concentration of radio electrons for different members of zone 1.

In Paper 1 we have shown that a high contrast of electron densities 
needed for interpretation of the radio observations can be 
reached if the zone 1 components correspond to the sites of efficient 
electron acceleration.  
In principle, the interpretation of the radio data of Cas~A does not suggest 
an efficient acceleration of the electrons in zone 2. We note however 
that such an acceleration, with similarly hard power-law index 
$\beta_{\rm acc}$ as in zone 1, is not excluded, provided that the amount 
of radio electrons accelerated directly in the plateau region does not 
significantly exceed the number of electrons leaking into zone 2 from 
zone 1. Otherwise the gradients in the spatial densities of electrons 
would be reduced resulting then in a strong
suppression of spectral modifications of $N_1(E,t)$ in the zone 1 regions. 

 The equation for the electron 
distribution $N_2(E,t)$ in zone 2 is similar to Eq.(1), where  
the source function $Q$ is substituted by $Q_{2}= N_1/\tau_{\rm esc}$
(taking into account that effectively the escape of relativistic 
electrons from zone 1 corresponds to their injection into zone 2, and 
neglecting possible acceleration in zone 2), and the 
sign of the term $V_1 N_2/V_2 \tau_{\rm dif}$ which describes 
the diffusive escape of electrons from zone 2 into zone 1 is changed.
Note that a two-zone model does not assume any other escape of electrons
from zone 2, e.g. into regions outside of the shell,
because otherwise such an escape would correspond to a three-zone model
(see Sect.~3 below).

Numerical calculations are done by the method of iterations using
general analytical solutions to the Leaky-box type equations (e.g. see 
Atoyan \& Aharonian 1999). For example, the solution for Eq.(1) can be 
presented as 
 \begin{eqnarray}
N_{1}(E,t) & = & \frac{1}{P_1(E)} \int_{0}^{t}
P_{1}(\zeta_t) Q_{\rm eff}(\zeta_t,t_1) \times \nonumber \\
& & \exp \left( -\int_{t_1}^{t}\frac{{\rm d} x}
{\tau_{\rm esc}(\zeta_x)}\right)
{\rm d} t_1\, ,
\end{eqnarray}
where $Q_{\rm eff}(E,t)=Q+N_2/\tau_{\rm esc}$. 
The  variable $\zeta_t$ corresponds to the energy of an electron at 
instant $t_1\leq t$ which has the energy $E$ at the 
instant $t$, and is determined from the equation
\begin{equation}
t-t_1 =\int_{E_{\rm e}}^{\zeta_t}{\frac{{\rm d}E}{P_1(E)}} \, .
\end{equation}
 At the first step of the iteration procedure a  
function $N_{1}^{(1)}(E)$ corresponding to the first approximation
of $N_1$ is calculated for an injection function $Q$ in the 
form of Eq.(4), i.e. neglecting the term $N_2/\tau_{\rm esc} $ 
in $Q_{\rm eff}(E)$, or else - neglecting in Eq.(1) the term
describing the diffusive `return' flux of the electrons from zone 2 into
zone 1. Then the energy distribution of the electrons $N_{2}^{(1)}(E)$ in
zone 2 in the first approximation, corresponding to
an effective injection function of the electrons 
in this zone $N_{1}^{(1)}(E)/\tau_{\rm esc}(E)$, is found. 
After that $N_{2}^{(1)}(E)$ is used for calculations of $N_{1}^{(2)}(E)$ 
in zone 1 in the second approximation, which now takes into account 
the `return' 
flux of the electrons from zone 2. Calculations show that such a simple 
iteration procedure is quickly converging, so typically several iterations
are sufficient to reach a good accuracy in the final electron 
distributions $N_1$ and $N_2$.   

%f1
\begin{figure}[htbp]
\resizebox{8.8cm}{!}{\includegraphics{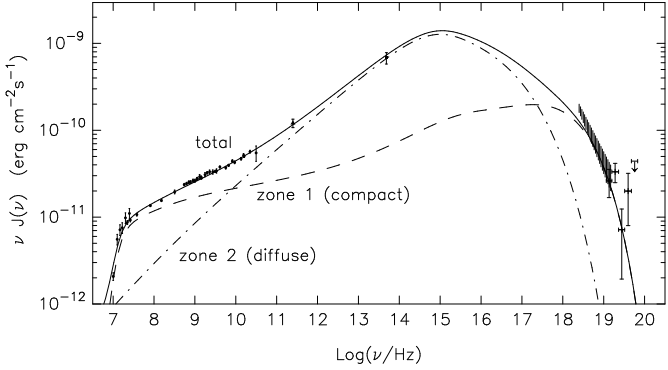}}
\caption{Synchrotron fluxes calculated in the framework 
of two-zone model for Cas~A. An acceleration spectrum of the 
electrons in the zone 1, which is modelled as being composed
of $K= 150$ compact structures with a mean radius $R=0.06\,\rm pc$
is given by Eq.(4) with the power law index 
$\beta_{\rm acc}=2.2$ and  exponential cutoff energy $E_{\rm c}=
18\,\rm TeV$;   escape of the electrons into the surrounding extended zone 2 
is described by Eqs.~(2) and (3), with 
parameters $\tau_{\ast}=28\,\rm yr$ and $\delta =0.7$. The 
magnetic fields in  zones 1 and 2 are $B_1 = 1.2\,\rm mG$ and 
$B_2=0.3\,\rm mG$, respectively. Dashed and dot-dashed curves
show contributions of zone 1 and zone 2 into the overall synchrotron
radiation flux (solid). The fluxes of Cas~A measured from radio to 
soft $\gamma$-ray bands (see Paper 1 for relevant 
references) are also shown. }
\end{figure}

%f2
\begin{figure}[htbp]
\hspace{5mm}\resizebox{7.5cm}{!}{\includegraphics{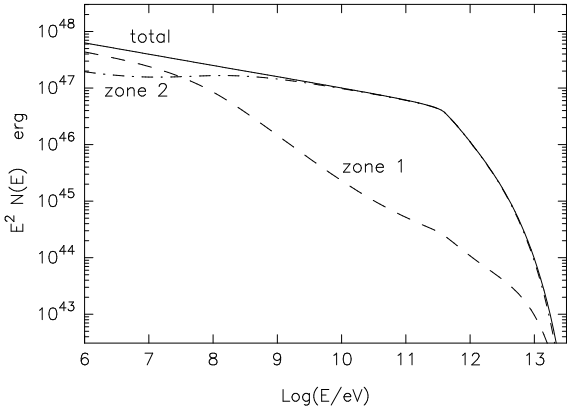}}
\caption{The energy distributions of relativistic electrons 
$N_1(E)$ and $N_2(E)$ in zone 1 (dashed line) and zone 2 (dot-dashed line)
which correspond to Fig.1. The solid line shows the total energy distribution
of electrons in the source.}
\end{figure}

The characteristics of the broad band 
synchrotron radiation of 
Cas~A can be best explained for the mean magnetic fields 
$B_{1}\simeq (1-2)\,\rm  mG$ and $B_{2}\simeq (0.3-0.5)\,\rm mG$ in zones 1 
and 2, respectively.    
The total energy content in relativistic electrons in each of these zones is  
of order $10^{48}\,\rm erg$, depending on the magnetic 
fields $B_{1}$ and $B_{2}$. The X-ray fluxes observed above $10\,\rm keV$ 
can be explained by synchrotron radiation of electrons in zone 1 with an  
exponential cutoff energy $E_{\rm c}\simeq (15-30)\,\rm TeV$ in Eq.(4).

A possible fit to the broad band synchrotron fluxes of Cas~A is shown 
in Fig.\,1. 
Magnetic fields $B_1=1.2\,\rm mG$ and $B_2=0.3\,\rm mG$ 
in the compact bright radio structures and in the diffuse plateau region, 
respectively, are 
assumed. The total magnetic field energy 
in the shell is then $W_{\rm B2}=3.8\times 10^{48}\,\rm erg$, 
and in the compact
zone 1 regions it is $W_{\rm B1}=2\times 10^{47}\,\rm erg$.
The total energy of electrons in zones 1 and 2 is  
$W_{\rm e1}=1.6\times 10^{48}\,\rm erg$ and  
 $W_{\rm e2}=1.8\times 10^{48}\,\rm erg$, respectively.

The energy distributions of the electrons $N_1(E)$ and $N_2(E)$ formed in 
 zones 1 and 2, as well as the overall distribution 
$N_{\rm tot}(E) = N_1(E)+N_2(E)$, are shown in Fig.\,2. As 
expected, $N_{\rm tot}(E) $ shows a pure power-law behavior with $\beta=
\beta_{\rm acc}$ until the so-called `radiative break' 
energy $\simeq 500\,\rm GeV$.  At this energy  
the synchrotron cooling time of the electrons in zone 2  
\begin{equation}
t_{\rm s}\approx 
300 \, (E /1{\,\rm TeV})^{-1} (B_2/0.2\,\rm mG)^{-2}\, yr
\end{equation}
becomes equal to the age of the source, and at higher energies 
the radiative losses steepen 
the spectrum of zone 2 electrons to the power-law 
index $\beta=\beta_{\rm acc}+1$.
Meanwhile, at $E\geq 100\,\rm MeV$ the energy distribution of radio 
electrons in zone 1 is steepened to $\beta_1\simeq 
\beta_{\rm acc}+\delta$  because of diffusive
escape of the electrons from that zone where their spatial density
$n_1(E)=N_1(E)/V_1$ is much higher than 
$n_2(E)=N_2(E)/V_2$ in the surrounding plateau region.

\subsection{Gamma-ray emission}

The photon flux $I(E)=J(E)/E$ of the bremsstrahlung $\gamma$-rays 
at energies $E\gg m_{\rm e}c^2$ practically repeats 
the power-law spectral shape  of the parent electrons. 
Therefore, one might  expect that $I(E)$ 
would have the same power-law index 
$\alpha_{\rm dif}=\alpha+1 \approx \beta_{\rm acc}$ as $N_{\rm tot}(E)$. 
However because of the different spectral shapes of the radio electrons in
zones 1 and 2 where, most probably, the gas densities are also different,
the total bremsstrahlung spectrum of Cas~A can somewhat deviate from 
the power-law behavior of the injection spectrum. 

%f3
\begin{figure}[htbp]
\resizebox{8.8cm}{!}{\includegraphics{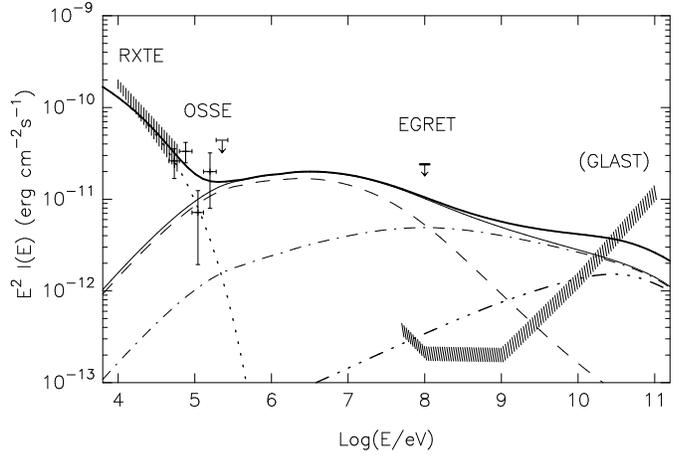}}
\caption{The fluxes of synchrotron (dotted line), bremsstrahlung (thin solid line)
and inverse Compton (3-dot--dashed line) radiations calculated in the framework
of two-zone  model for the same model parameters as in Fig.~1. The dashed 
and dot-dashed curves correspond to the bremsstrahlung fluxes produced
in zone 1 and zone 2, respectively. The heavy solid curve is the overall
flux. The hatched region shows  the expected flux sensitivity of  GLAST
(Bloom 1996). The X-ray/soft $\gamma$-ray fluxes measured 
by RXTE (Allen et al. 1997) and OSSE (The et al. 1996) detectors, as well as flux
upper limit of EGRET (Esposito et al. 1996) are also shown.} 
\end{figure}

In Fig.3 the thin solid line corresponds to the total 
flux of the bremsstrahlung photons produced by the electrons shown in Fig.2, 
and the dashed and dot-dashed lines represent the fluxes from  zone 1 
and zone 2, respectively.
For the calculations we adopt  a mean 
gas density $n_{\rm H,2}=15 \,\rm cm^{-3}$ in terms of `H-atoms' 
(i.e. the nucleons) in zone 2. This corresponds to a total 
mass of about $15\,M_\odot$ (e.g. Fabian et al. 1980; Jansen et al. 1988; 
Vink et al. 1998), in the volume $V\approx 1.2\times 10^{57}\,\rm cm^3$ 
of the shell between $R_{\rm ring}$ and $R_0$. 
The mean value of the parameter $C_{\rm Z}=Z (Z+1)/A$ in the oxygen-rich 
($Z=8,\; A=16$) gas in the shell (zone 2) of Cas~A derived by 
Cowsik \& Sarkar (1980) is $\overline{C_{\rm Z}} = 4.3$.
For zone 1 we assume the same atomic $\overline{C_{\rm Z,1}}=4.3$, and 
the gas density $n_{\rm H,1}=4 n_{\rm H,2}$.
We note however that these parameters in the compact radio structures 
are not known. In particular, the radio knots 
show practically no optical line emission which would allow conclusions 
about the density and mass composition of the gas there. Therefore,
for different values of the gas parameters in zone 1, the relative contribution 
of zones 1 and 2 to the total bremsstrahlung flux would change. As shown in
Fig.4, this results in some uncertainty of the model predictions for the 
steepness of the total $\gamma$-ray fluxes at energies   
$E\sim 100\,\rm MeV$. At both smaller ($\leq 10\,\rm MeV$)
and higher ($\geq 1\,\rm GeV$) energies, where the bremsstrahlung flux is  
contributed mainly either by the first or by the second zone, 
the spectral indices are much less affected
by the uncertainty in the gas parameters in zone 1.

Because of the steep decline of the energy distribution of radio electrons
in the compact zone 1 structures, the intensity of the $\gamma$-ray flux 
at $E\sim 1\,\rm GeV$ is dominated by the flat-spectrum bremsstrahlung
of zone 2 (see Fig.3). It is also important that the contribution of
the IC radiation, discussed in detail in the next section, at this energy
is not yet significant. Therefore,  measurements of the 
differential flux $I(E)$ of high energy $\gamma$-rays 
from Cas~A by the future GLAST detector should allow a rather accurate 
determinations of a number of important parameters in Cas~A.
In particular,  
the spectral index of $I(E)$ at GeV energies could give rather
accurate information about the spectrum of electrons in zone 2,
$\alpha_{\rm dif}\approx \beta_2 $ (see Figs. 3 and 4). At these energies 
the energy distribution of  electrons in zone 2 has a 
power law distribution $N_{2}(E)\propto E^{-\beta_2}$ with the spectral 
index of accelerated particles, $\beta_2= \beta_{\rm acc}$ (see Fig.2). 

%f4
\begin{figure}[t]
\hspace{5mm} \resizebox{7.2cm}{!}{\includegraphics{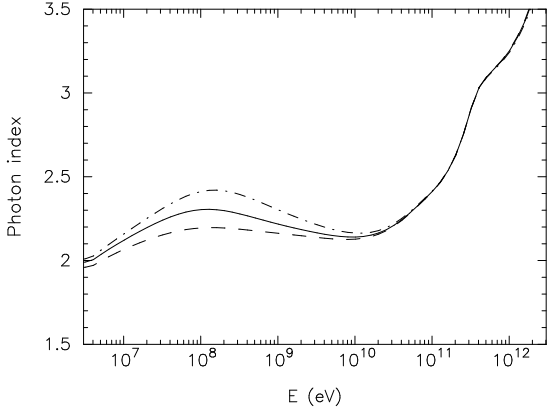}}
\caption{The spectral indices of the photon flux $I(E)$ of the broad-band 
$\gamma$-radiation expected from Cas~A in the case of 3 different ratios
of the gas parameter $n_{\rm H}\,\overline{C_{\rm Z}}$ in the compact 
zone 1 structures and extended zone 2 (the shell). The solid curve corresponds
to the ratio $  n_{\rm H,1}\,\overline{C_{\rm Z,1}}/  
n_{\rm H,2}\,\overline{C_{\rm Z,2}} = 4$, as assumed for the calculations in
Fig.~3, whereas the dashed and dot-dashed curves show the spectral photon
indices which result if this ratio is equal, respectively,  to 2 and 8.}
\end{figure}

Bremsstrahlung $\gamma$-rays with energies $E_{\gamma}\simeq 1\,\rm GeV$ are  
produced by  electrons with characteristic energies $E_{\rm e}\sim 2\,E_{\gamma}$.
The mean frequency of the synchrotron photons emitted by the same electrons
is $\nu  \simeq 10^3 B_{\rm mG} \, (E_{\rm e}/m_{\rm e}c^2)^2$ 
where $B_{\rm mG} = B/1\,\rm mG$ (e.g. Ginzburg 1979).  
In terms  of the $\gamma$-ray energy this relation is reduced to 
\begin{equation}
\nu  \simeq 4\, B_{\rm mG} \, (E_\gamma / \rm 1\, GeV)^2 \; GHz\, .
\end{equation}
For the expected mean magnetic field in the shell of Cas~A of about 
(0.3--0.5)\,mG (Paper 1),  electrons responsible for bremsstrahlung   
of (1-2)\,GeV $\gamma$-rays are also producing  
synchrotron photons with $\nu \sim$ few GHz. Thus, a comparison of the 
radiation intensities in these two regions  will give almost model 
independent information about the mean magnetic field $B_2$ in the shell. 

Indeed, for the power-law distribution of electrons $N(E)\propto 
E^{-\beta}$, the intensity of synchrotron radiation can be written
in the form (see Paper 1):
\begin{eqnarray}
J_\nu &\simeq &1.5\times 10^{3}\left(\frac{ N_{\ast}}{10^{50}}\right) 
\left( \frac{B}{1\,\rm mG} \right)^{\frac{1+\beta}{2}}\times \nonumber \\
& & \left( \frac{\nu}{10\,{\rm GHz}}\right)^{\frac{1-\beta}{2}}
\left(\frac{d}{3.4\,\rm kpc}\right)^{-2} \;
\rm Jy\; ,
\end{eqnarray}
where $N_{\ast} \equiv \, E_{\ast}\,N(E_{\ast})$, with $E_\ast =1\,\rm GeV$,
is the characteristic total number of 1\,GeV electrons. 
The  bremsstrahlung intensity I(E) 
(see e.g. Blumenthal and Gould 1970) can be presented  
  in the form of an energy flux $f(E)=E^2 I(E)$ as 
\begin{eqnarray}
f(E) & \simeq & 7.5\times 10^{-15} \,  n_{\rm H}\,\overline{C_{\rm Z}}\,
 2^{2-\beta} \left(\frac{ N_{\ast}}{10^{50}}\right) 
 \left(\frac{E}{1\,\rm GeV}\right)^{2-\beta} 
 \nonumber \\
& &
\times \left[ \ln \left(\frac{E}{1\,\rm GeV} \right) +8.4 \right]  
\left(\frac{d}{3.4\,\rm kpc}\right)^{-2}
\, \rm 
\frac{erg}{cm^2 s}\, ,
\end{eqnarray}
where $\overline{C_{\rm Z}}=\overline{Z (Z+1)/A}$, as previously. Then, because 
the total intensity of $\gamma$-rays at (1-2) GeV is dominated by the 
bremsstrahlung of radio electrons in the shell (zone 2) responsible also for 
the diffuse `radio plateau' emission
at $\nu \sim 5\,\rm GHz$ with the known flux $J_2(\nu)$ \footnote{ The 
measurements of Tuffs (1986) taken in 1978 have shown that $\approx 50\,\%$ 
of the total flux $J(\nu)
\approx 800\,\rm Jy$ of Cas~A was due to plateau emission. Taken into
 account the  
secular decline of the fluxes, 
the intensity of diffuse radio emission presently can be
estimated as  $J_2(5\,\rm GHz)\sim 350\, Jy$.},  and because the value of 
$n_{\rm H,2} \,\overline{C_{\rm Z,2}}$ in the shell is known with sufficient 
accuracy, comparison of  equations (9) and (10) will result in a  rather 
accurate estimate of the 
magnetic field $B_2$. For example, in the case of $\alpha_{\rm dif}\simeq 
\beta_{\rm acc} \sim 2.2$ as expected, the mean magnetic field in the shell
would be   
\begin{equation}
B_2\approx 0.16 \left[ \frac{J_2(5\,\rm GHz)}{350\,\rm Jy}\right]^{0.63}
\left[\frac{f(1\,\rm GeV)}{10^{-11}\,\rm erg/cm^2 s}\right]^{-0.63} \rm mG 
\end{equation}
This knowledge of $B_2$ could then help to estimate the mean magnetic field
also in compact bright radio structures as $B_1\sim 4 \,B_2$ which is 
needed for interpretation of the radio data (see Paper 1). 

At energies $\leq 100\,\rm MeV$ the overall flux of $\gamma$-rays is 
dominated by the bremsstrahlung of relativistic 
electrons in the compact zone 1 region. As follows 
from Eq.(8), the radio counterpart of the $E\sim 100\,\rm MeV$ $\gamma$-rays
is the region of frequencies $\nu \sim 40\,\rm MHz$, where the synchrotron 
radiation is dominated by zone 1 emission (see Fig.1), and the fluxes are not
yet affected by synchrotron self-absorption. Although the expected flux sensitivity 
of the GLAST detector drops significantly 
below 100\,MeV, in principle 
it may be able to detect $\gamma$-ray fluxes down to energies 
$\sim 10\,\rm MeV$ (Bloom 1996).
Detailed  modelling of the flux $I(E)$  would help to extract the intensity
of 100\, MeV bremsstrahlung $\gamma$-rays  produced in zone 1,
i.e. in the bright radio ring and radio knots. Then, 
for known  $I_1(E\sim 100\,\rm MeV)$, $J_1(\nu = 40\,\rm MHz$), and 
$\beta = \beta_1\sim (2.7-2.8)$ in zone 1 (as predicted in Paper 1), 
the equations (9) and (10) can be used to derive  the unknown product  
$(n_{\rm H} \times \overline{C_{\rm Z }})$ there. 
Note that in principle the gas parameters
can be different in the radio ring and radio knots. 
A possibility of separation of the
contributions of these two different sub-components of zone 1 to the overall flux  
$I_1(E)$ could enable determination of the product   
$n_{\rm H} \overline{C_{\rm Z }}$ in each of them. 

Important additional information about the gas parameters in 
compact components could be derived from   
future observations of Cas~A with  high angular and energy resolution 
by the  X-ray  
telescopes Chandra, XMM and Astro-E  in the keV region. 
Even for the case of non-detection of the line  
features from radio knots, the intensity of their  
{\it thermal} X-ray emission will help to disentangle the 
density $n_{\rm H}$ from the 
mean abundance $ \overline{C_{\rm Z }}$. Because the intensity of thermal 
emission $Q_{\rm T}\propto  \overline{Z^2} \,n_{\rm Z} n_{\rm e}\simeq 
\overline{C_{\rm Z}} n_{\rm H} n_{\rm e}$, and because for the ionized  gas the 
thermal electron density $n_{\rm e}= Z n_{\rm Z}\simeq n_{\rm H}/2 $ 
(except for a 
hydrogen-dominated medium where $n_{\rm e}\simeq n_{\rm H}$), the 
fluxes of thermal
X-rays will in principle allow determination of the parameter 
$\overline{C_{\rm Z}} n_{\rm H}^2$ in the zone 1 structures. When combined with
the knowledge of a different product of parameters $\overline{C_{\rm Z}}$ 
and $n_{\rm H}$ found from the comparison of  $\gamma$-ray with  
synchrotron measurements, each of these two parameters could be found.

\section{Gamma-ray emission at very high energies}

\subsection{Emission of relativistic electrons} 

If the fluxes of hard X-rays observed at $E\geq 10\,\rm keV$ have a 
synchrotron origin (Allen et al. 1997, Favata et al. 1997; -- but see also
Laming 1998), relativistic electrons in Cas~A 
should be accelerated to energies up to tens of TeV. These electrons 
should then
produce very high energy (VHE, $E\geq 100\,\rm GeV$) $\gamma$-rays.
Along with bremsstrahlung, at these energies the principal mechanism for 
$\gamma$-ray production is the inverse Compton (IC) scattering of 
the electrons 
in the ambient soft photon field. In principle, the photon field is 
contributed 
by the synchrotron photons, the thermal dust emission with $T=97\,\rm K$
(Mezger et al. 1986) in the far infrared (FIR), 
the optical/IR line photons, and the 2.7\,K cosmic microwave background
radiation.

As shown in Fig.5, the most important target photon field for 
production of IC $\gamma$-rays  
in Cas~A is the FIR radiation which is responsible for $\geq 80\%$
of the IC flux in the VHE region. Note that only due to the high 
density of the FIR radiation in Cas~A (which has not been taken
into account in recent calculations by Ellison et al. 1999), the flux of 
IC $\gamma$-rays becomes comparable at TeV energies with the bremsstrahlung 
flux. 
Because the photon fields should have practically 
the same density in both compact and diffuse zones, irrespective of
where they are produced, and since the VHE electrons 
reside mostly in zone 2 (see Fig. 2), IC radiation is contributed
mostly by this zone. For an assumed 
mean magnetic field in the shell 
$B_2 = 0.3\,\rm mG$ (Fig.5), the energy density of the 
magnetic field is $w_{\rm B}=
2.5 \times 10^3 \,\rm eV/cm^3$, whereas the density of the FIR radiation 
calculated for the  luminosity $L_{\rm FIR}\approx 3.6\times 10^{37}\,
\rm erg/s$ (Mezger et al 1986) is only $w_{\rm rad} \simeq 2\,\rm eV/cm^3$. 
Because the ratio of synchrotron to IC (in the Thomson limit)
emissivities
\begin{equation}
q_{\rm sy}/q_{\rm IC}= w_{\rm B}/ w_{\rm rad} \propto B^2 \;,
\end{equation}
the IC fluxes of VHE $\gamma$-rays 
are $\simeq 10^3$ times lower than the fluxes of synchrotron radiation 
produced by the
same parent electrons in the UV/X-ray  region.  
       
%f5
\begin{figure}[htbp]
\resizebox{8.5cm}{!}{\includegraphics{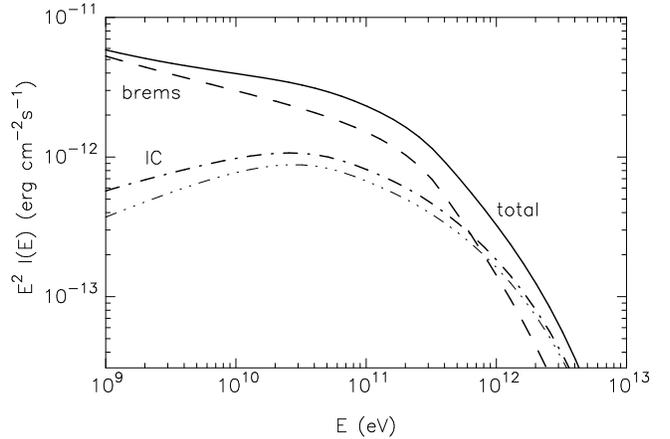}}
\caption{The fluxes of bremsstrahlung (dashed) and IC (dot-dashed)
$\gamma$-rays calculated in the framework of two-zone model 
for Cas~A for the same model parameters as in Fig.1. The solid curve
corresponds to the total flux. The thin 3-dot--dashed curve
shows the fraction of IC $\gamma$-rays produced due to upscattering of 
the thermal dust photons; other target photons for IC scattering of electrons
are the synchrotron radiation, 2.7\,K microwave background radiation,
and optical line emission of Cas~A.}   
\end{figure}

The fluxes of  IC $\gamma$-rays to be expected from Cas~A depend very 
sensitively on the mean magnetic field $B_2$ in the shell. 
As follows from Eq.(9), an 
increase of 
$B_2$ by a factor $a$ reduces the total number of electrons needed for
explanation of the radio fluxes observed by a factor $a^{(1+\beta_2)/2}$. 
This results in a strong decrease of the bremsstrahlung flux by the same factor,
i.e. $f_{\rm br} \propto 1/a^{(1+\beta_2)/2}$. The same dependence on the
magnetic field also holds for the intensity of IC emission which is
produced by the electrons below the radiative break 
energy (the `knee'
around 500 GeV in Fig.2) for which the synchrotron cooling time 
is larger than the age $t_0$ of the source. For these electrons 
the energy distribution $N(E)\approx Q(E)\times t_0$ repeats the injection 
spectrum $Q(E)$. However, VHE $\gamma$-rays are produced by electrons of  
higher energies. In that case the spectrum of electrons is 
$N(E)\approx Q(E)\times t_{\rm s}$. Then, besides of a less powerful
injection rate $Q\propto a^{-(1+\beta_2)/2}$ needed for explanation of the
radio fluxes, the spectral intensity of these electrons is additionally 
suppressed by a factor $a^2\propto B_{2}^2$. Therefore the intensity
of VHE radiation depends  
on the magnetic field in zone 2 as $f\propto B_{2}^{-(5+\beta_2)/2}$, 
which is significantly stronger than at GeV energies. 
This is the result of emission 
of VHE electrons in the ``saturation'' regime, when all the power
injected in TeV electrons is channeled into (mostly) synchrotron
and IC fluxes in the proportion defined by Eq.(12).

Thus, the flux of IC $\gamma$-rays  could be significantly 
increased assuming {\it smaller} values of $B_2$,  and hence also of $B_1$ 
because the ratio of magnetic fields in two zones should be approximately 
at the level $B_1/B_2\sim 4$ in order to explain the radio data 
(see Paper 1). On the other
hand, $B_2$  cannot be significantly less than 0.3\,mG, 
otherwise the bremsstrahlung flux would then exceed the radiation flux observed
at $E\sim 100\,\rm keV$, and the flux upper limit of 
EGRET at $E\geq 100\,\rm MeV$
(see Fig. 3). Note that this constraint on the magnetic field in the shell
of Cas~A imposed
by the EGRET data are model 
independent since  $E\geq 100\,\rm MeV$ $\gamma$-rays are produced by
the same electrons with energies $E_{\rm e} \geq 1\,\rm GeV$ 
which are responsible for the observed radio fluxes. On the 
other hand, the constraints imposed by the hard X-ray fluxes are to some 
extent conditional, being based on a (resonable) assumption that the power-law
injection spectrum starts from sub-MeV energies.

 It is worthwhile to compare the constraints for the mean
magnetic field in the shell, i.e. $B_2\geq 0.3 \,\rm mG$, with the 
constraints imposed in the framework of a spatially homogeneous 
(i.e. a single-zone) model commonly used. 
Because radiative losses cannot steepen the spectrum of the radio electrons
in Cas~A,
both the acceleration spectrum and the overall energy
distribution of electrons $N(E)$ at GeV energies should be steep, with a 
power-law index $\beta_{\rm acc}\geq 2.5$ for the mean power-law index of
the observed radio fluxes $\alpha\simeq 0.77$. This is obviously much 
steeper than the index $\beta_{\rm acc}\simeq 2.2$ predicted by the two-zone
model. As a result, the 
overall number of electrons predicted in the framework of such
single-zone approach at low energies is much higher than for the two-zone
model, and hence the lower limit for the mean magnetic field in the shell 
of Cas~A consistent with the X-ray fluxes and upper flux limit  
increases to $B_{\rm min} \sim 1\,\rm mG$ (e.g. Ellison et al. 1999). 
Both the steepness of the spectrum and the higher magnetic field predicted
in the framework of a single-zone model significantly reduce the  
fluxes of $\gamma$-rays to be expected at very high energies. 
Note that the impact of the steepness of the acceleration spectrum 
$Q(E)$ can be significantly reduced in the framework of a 
more elaborated model where $Q(E)$ becomes significantly flatter at 
higher electron energies, as in calculations by  
Ellison et al. (1999). However, there is no way to reduce the
impact of the high mean magnetic field 
$\geq 1\,\rm mG$ which essentially reduces the number of multi-TeV
electrons needed for production of a given flux of synchrotron
X-rays. Therefore,  in the framework of a single-zone model 
the fluxes of VHE radiation which would {\it not contradict} the 
observed hard X-ray flux will be significantly lower than
the fluxes to be expected  in the framework of a spatially
inhomogeneous two-zone model (compare Fig.~5 with the results
of Ellison et al. 1999, and Goret et al. 1999).

For the same values of the mean magnetic fields in zones 1 and 2, the 
fluxes of IC  $\gamma$-rays expected from Cas~A at TeV energies could be 
higher than in Fig.\,5 if we consider a more structured model for the magnetic 
field distribution in the shell than the 
two-zone model. Namely, we may assume that the magnetic field
in the shell of Cas~A may decrease from the highest value $B_1$ in the compact
zone 1 (the acceleration sites) to a lower value $B_2$ in the surrounding 
zone 2, and further on to some $B_3 \le B_2$ in zone 3. The chain of 
equations describing the energy distributions of  particles in such 
3-zone model is easily derived in the
same way as described in Paper 1 for the 2-zone case. Note that the 3-zone
modelling may be more adequate to the radio pattern  
 of Cas~A which shows significant variations in the brightness of the 
diffuse emission of the shell. Zone 3 would then represent the regions of the
shell with relatively low magnetic field, as well as possibly the regions
adjacent to the shell. Relativistic electrons could then escape from zone 2
into zone 3, as they do from zone 1 into zone 2.
Because the spatial sizes
of zone 2 are significantly larger than the sizes of the compact zone 1 
structures, the characteristic timescales for the electron escape from 
zone 2 into zone 3 should be significantly larger than that 
the escape time from zone 1 into zone 2.

In Fig.6 we show the synchrotron radiation spectra calculated in the framework 
of the 3-zone model assuming that zones 2 and 3 have 
approximately the same volume filling factors in the shell. Note that
zone 3 may include also regions not contained by the shell, in particular, the 
volume interior to the reverse shock if the diffusion coefficient 
there is sufficiently high (e.g. because of a low magnetic field there)
so that relativistic particles would be able to significantly 
penetrate upstream of the freely
expanding ejecta. In any case,  
zone 3 gives a principal possibility for high energy electrons to escape from 
zone 2 into a region with a lower magnetic field, resulting in some steepening of
$N_2(E)$ at $E\geq 10\,\rm GeV$. Then  it becomes possible to
assume a power-law injection spectrum of the accelerated particles (in zone 1) 
harder than $\beta_{\rm acc}\approx 2.2$,  
without an excess of the radiation fluxes measured at $6\,\rm \mu m$ 
(Tuffs et al., 1997). 

%fig6
\begin{figure}[htbp]
\resizebox{8.7cm}{!}{\includegraphics{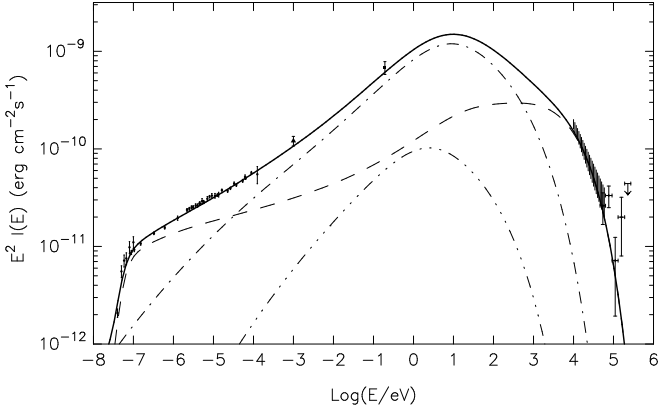}}
\caption{The fluxes of synchrotron radiation  calculated in the framework of
the three-zone model, when the shell may contain regions with higher (zone 2)
and lower (zone 3) magnetic fileds. The fluxes produced in zones 1, 2 and 3,
with the assumed magnetic fileds $B_1=1.6\,\rm mG$, $B_2=0.4\,\rm mG$
and $B_3=0.1 \,\rm mG$, are plotted in dashed, dot-dashed, and 
3-dot--dashed lines, respectively. The heavy solid line shows the total flux.
The injection spectrum of the electrons in zone 1 is in the form of Eq.(4)
with $\beta_{\rm acc}=2.15$ and $E_{\rm c}=17\,\rm TeV$; the 
diffusive escape  of electrons  is described by $\delta=0.7$ for both zones 1 
and 2, but $\tau_{\ast, 1}=20\,\rm yr$ for compact zone 1, and 
$\tau_{\ast, 2}=800\,\rm yr$ for the much larger zone 2.}
\end{figure}

%fig7
\begin{figure}[htbp]
\resizebox{8.5cm}{!}{\includegraphics{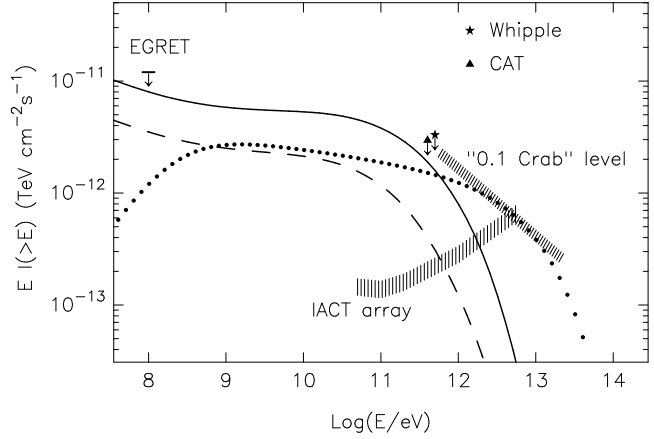}}
\caption{The integral fluxes of the broad-band $\gamma$-radiation
expected from Cas~A in the case of two different values of the magnetic
field in zone 1, 
 $B_1=1\,\rm mG$ (solid line) and $B_1=1.6\,\rm mG$ (dashed line),
calculated in the framework of a 3-zone model, assuming that $B_2=B_1/4$ 
in zone 2, $B_3=0.1\,\rm mG$, and an injection of relativistic electrons 
with $\beta_{\rm acc}= 2.15$. 
All other model parameters are the same as in Fig.~6, except for 
the exponential cutoff energy in the case of $B_1=1\,\rm mG$ for which
$E_{\rm c}=12\,\rm TeV$ (in order to fit the X-ray fluxes for this magnetic
field). The full dots show the fluxes of $\pi^0$-decay $\gamma$-rays calculated
for relativistic protons with total energy $W_{\rm p}=2\times 10^{49}\,\rm erg$,
and acceleration spectrum with $\beta_{\rm pr}=2.15$ and $E_{\rm c}=200
\,\rm TeV$, for a typical `H-atom' gas density in the shell of Cas~A about
$n_{\rm H}=15\,\rm cm^{-3}$. The level of fluxes corresponding to $10\,\%$
of the flux detected from the Crab Nebula (e.g. Konopelko et al 1999),
and the expected range of flux sensitivities of future IACT arrays  
for $n_0 t= 10^3\,\rm h$ (where $n_0$ is the number of cells, see Aharonian 
et al. 1997) are also shown.}  
\end{figure}

In Fig. 7 we show the integral fluxes  of IC $\gamma$-rays, in terms
of $E\times I(>E)$, calculated in the framework
of the 3-zone model, assuming 2 different values for the magnetic
field in zone 1, $B_1=1\,\rm m G$ (solid line) and $B_1=1.6\,\rm m G$ (dashed
line), the magnetic field $B_2=B_1/4$ in zone 2, and $B_3=0.1\,\rm mG$ in
zone 3. Although $B_1$ and $B_2$ in
these 2 cases change only by a factor 1.6, the fluxes of TeV $\gamma$-rays 
drop dramatically, by a factor 
of 6-7.  
Note that an assumption of the magnetic field $B_3$ for zone 3 
smaller than 
0.1\,mG does not result in a further increase  of  TeV $\gamma$-ray fluxes, 
because for such low magnetic fields all electrons up to several TeV are not
in the ``saturation'' regime (see Eq.7), therefore
variations of $B_3$ do not affect the electron energy distribution $N_3(E)$.

 It should be noted in connection with the 3-zone model, as compared 
with the 2-zone model,that it allows an increase 
of the $\gamma$-ray fluxes at energies above 1 TeV (where the IC
radiation dominates the overall flux from electrons) by a factor of 3 assuming
the same magnetic fields $B_1$ and $B_2$ for zones 1 and 2.
Because the model parameters in zone 3 (as the magnetic field 
$B_3$, the volume $V_3$ of zone 3 with this low field, and the escape time
from zone 2 into zone 3) are difficult to deduce from  
radio observations, this introduces an additional uncertainty 
in the model predictions for  the fluxes of VHE radiation. 
However, the mean magnetic fields $B_2$ and $B_1$ can be deduced rather
accurately from future $\gamma$-ray detections
of Cas~A at lower energies (where bremsstrahlung dominates) as discussed 
in Sect.~2\,. Then in principle the   
spectral measurements of VHE $\gamma$-ray fluxes could give an important
information about the parameters of zone 3 with low magneic field.

The fluxes of TeV $\gamma$-rays on the level down to few per cent of
the  Crab nebula flux are in 
principle accessible for a system of Imaging 
Atmospheric Cherenkov Telescopes (IACTs) like the presently operating
HEGRA. In this respect, the recent report of the HEGRA collaboration 
(P\"uhlhofer et al. 1999) about
possible detection, at $\sim 5\sigma$ significance level, of a weak
signal from Cas~A appears very interesting, and needs a further 
confirmation.   
It should be said that the extraction of a signal from Cas~A 
would require special care, because the fluxes of IC $\gamma$-rays 
are expected to  decline already at energies 
$E\geq 0.5\,\rm TeV$ much faster than the flux of the Crab Nebula,  
which is generally considered as a ``standard candle''. This can be seen
in Fig.~7, as well as in Fig.~4 where the spectral index of the 
differential flux 
is plotted (to be compared with $\alpha_{\rm dif}\sim 1.5-1.6$ for the
Crab Nebula).
In Fig.~7 we also show the expected range of flux sensitivities of the
forthcoming IACT arrays. It predicts that a significant flux of 
VHE $\gamma$-rays 
should be observed by the future VERITAS array (which is to operate in 
the northern hemisphere), if the observed 
fluxes of hard X-rays above 10\,keV are indeed of synchrotron origin 
(Allen et al. 1997; Favata et al. 1997). Detection of the VHE $\gamma$-rays 
 would allow a rather robust estimate of the mean magnetic fields 
$B_1$ and $B_2$. This
could be done by a modelling, rather than by a direct comparison
of the X-ray and TeV $\gamma$-ray fluxes, because they are produced 
in different zones -- 
in compact zone 1 and the diffuse `radio plateau', respectively. 

\subsection{$\pi^0$-decay gamma-rays}

The  TeV $\gamma$-radiation in Cas~A could also be efficiently 
produced by relativistic protons and nuclei which should be
accelerated simultaneously with the electrons. Our study of the spectral and 
morphological characteristics of the radio emission of Cas~A does not 
favour the blast wave as an efficient accelerator of an electrons 
competitive with electron acceleration in the compact structures. However, 
there are no comparable
observational constraints on the proton acceleration sites. 
In particular, the protons could plausibly be accelerated also at the blast
wave. The standard theory of diffusive shock acceleration 
suggests that the hadronic
component of cosmic rays could be accelerated much more copiously than 
the electrons, which generally explains the up to two
orders of magnitude overabundance of the nucleonic component in the observed
galactic cosmic rays. 

In Fig.7 we show by full dots the $\gamma$-ray fluxes resulting 
from the decay of $\pi^0$-mesons produced by relativistic protons in the 
inelastic
interactions with surrounding gas in the shell of Cas~A. 
A power-law injection spectrum of relativistic protons in the form of Eq.(4), 
 with $\beta_{\rm acc}=2.15$ as for the electrons, is supposed.
 For  the characteristic maximum energy of accelerated protons we have 
assumed  $E_{\rm c}=200\,\rm TeV$. Given the very high magnetic field 
in the acceleration
region, $B_1 \geq 1\,\rm mG$, even for a rather  young age of Cas~A  
the relativistic protons, not being affected by synchrotron 
losses like the electrons, can easily reach such high values of $E_{\rm c}$.
Indeed, using an estimate for the standard `parallel shock' acceleration
efficiency, the characteristic maximum energy of the protons accelerated during
time $t_0$ (e.g. see Lagage \& Cesarsky, 1983) can be written as 
 \begin{eqnarray}
E_{\rm c} & \simeq & 450 \, \left( \frac{B}{1\, \rm m G} \right) \, \left( 
\frac{t_0}{100 \, \rm yr} \right) \times  \nonumber \\ 
& & \left( \frac{u_{\rm s}}{3000 \, \rm km/s} \right)^2 \, \eta^{-1} \, 
\rm TeV \, , 
\end{eqnarray} 
 where $u_{\rm s} \sim 3000\,\rm km/s$ is a typical shock speed in Cas~A 
(in particular, of the reverse shock), and $\eta\geq 1$ is the so called 
gyrofactor (the ratio of the mean free path of a particle to its gyroradius).
Thus, in the case of shock acceleration in the regime close to the 
Bohm diffusion limit, $\eta \sim 1$, the protons in Cas~A could reach the 
energy of 
order 500\, TeV during an acceleration time as short as $t_0\sim 100\,\rm yr$. 

The total  energy of relativistic protons assumed for the calculation of 
$\pi^0$-decay
$\gamma$-ray fluxes in Fig.~7 is  $W_{\rm p}=2\times 10^{49}\,\rm erg$,
 which corresponds to the mean injection power of the protons during
$t_0\simeq 300\,\rm yr$ of about $L_{\rm p}\simeq 2\times 10^{39}\,\rm erg/s$.
On the other hand, for the magnetic field $B_1=1\,\rm mG$ the acceleration 
power of the electrons is $L_{\rm e}=5.5\times 10^{38}\,\rm erg/s$, and
$L_{\rm e}=2.5\times 10^{38}\,\rm erg/s$ for  $B_1=1.6\,\rm mG$. Thus,
the protons are supposed to be accelerated only by a factor of $4-8$ more 
effectively than the electrons. This ratio cannot be increased further 
by more than a factor of 2, 
because otherwise the fluxes of TeV $\pi^0$-decay 
 $\gamma$-rays would exceed the flux  upper limits shown in Fig.~7.
Therefore the ratio of 
proton to electron acceleration efficiencies in Cas~A 
is significantly smaller than,  or else have not yet reached, 
the high value $\geq 40$ which is usually supposed for a typical 
source of the Galactic cosmic rays.
 
It is worth noticing that the upper limit to $W_{\rm p}$ imposed by 
the non-detection of TeV radiation from Cas~A by Whipple and CAT detectors, 
is by one order 
of magnitude lower than the limitation imposed by the flux upper limit
of the EGRET detector (see Fig. 7).
Most probably, the total energy in accelerated protons in Cas~A can be already
now limited by a value not significantly exceeding $10^{49}\,\rm erg$.
Of course, we cannot exclude  that relativistic 
protons are accelerated only to energies below the TeV range. This would
then imply that the electrons as well are not accelerated to these energies,
and therefore that the hard X-ray fluxes detected from Cas~A are not of a 
synchrotron origin (which cannot be still ruled out, see Laming 1998).
Thus, we expect that rather important conclusions
 could be derived already from the fact of, hopefully, further confirmation 
of the possible HEGRA detection of TeV $\gamma$-rays Cas~A.

For both hadronic and electronic 
origin of the TeV $\gamma$-rays, the fluxes are to be produced in the extended
shell of Cas~A enclosed between angular radii 1.5 and 2.5 arcmin, or perhaps even
in a bit larger region, if the VHE electrons would escape from the shell
(this may compose a part of zone 3, as discussed in previous section). In terms
of diameter, this makes an appreciable size of order of 5 arcmin. 
The IACT arrays are able to reconstruct the direction of individual  $\gamma$-rays 
with an accuracy of several arcminutes which, in combination with a significant 
statistics of the $\gamma$-rays, may result in a principal possibility to localize 
relatively strong point sources with an accuracy of about 1 arcmin 
(Aharonian et al.
1997). Therefore IACT arrays operating at 
energies $\sim 100\,\rm GeV$ might be able in principle to see some structure
(possible `hot spots') in the expected generally circular 
 VHE $\gamma$-ray image of the source.  

In this regard it might be worthwhile 
to note that radio observations of Kassim et al. (1995) at low frequencies
show definite signs of the presence of a substantial mass  (at least
several $M_\odot$)  of cold gas in 
the central
$\leq 1^\prime$ region of the Cas~A. If TeV particles would be able 
to penetrate
into that region (by diffusive propagation upstream of the freely
expanding ejecta -- in the case of high diffusion coefficient), then a central
 `hot spot' would appear, associated with the bremsstrahlung and $\pi^0$-decay
$\gamma$-rays produced by TeV electrons and protons illuminating the cold
unshocked ejecta.  The mean gas density of this ejecta in the central
$\leq 1^\prime $ region of Cas~A can be estimated as $n_{\rm H}\geq
10 \,\rm cm^{-3}$ per each solar mass of the ejecta (to be compared with
the mean $n_{\rm H}\simeq 15\,\rm cm^{-3}$ in the shell). Therefore, if
the multi-TeV particles could really diffuse into that region, so 
that, say, 5 per cent of the total amount of these particles (protons and/or
electrons) would be concentrated in that region, and the mass of freely
expanding ejecta there could be about $6\,\rm M_{\odot}$, then one could  
expect that up to $20\,\%$ of the total flux of VHE $\gamma$-rays 
would come from the centre of Cas~A.   
Note that at energies $\geq 100\,\rm GeV$ where the sensitivity of IACT arrays
is very high, most of the $\gamma$-ray flux produced by the electrons has
a bremsstrahlung origin (see Fig.5). 

Another possible `hot spot' of the IC origin could appear (at higher 
energies) because of the known
 asymmetric spatial distribution of the target IR photons near the spectral
peak around $\sim 30\,\rm \mu m$  
(see $25\,\rm \mu m$ map by Dwek et al. 1987). If TeV electrons 
are distributed 
rather homogeneously in the shell, then the expected overall IC flux would not
significantly change. However the image of Cas~A at TeV energies will shift
towards the position of maximum IR emission in the northern shell. If,
on the other hand, the density of TeV electrons were enhanced in the northern 
shell (where the radio brightness is also enhanced) then we could expect
an enhancement of the overall IC flux as well.

\section{Conclusions}

Although not yet conclusively 
detected, $\gamma$-rays should inevitably be  
produced in Cas~A by relativistic electrons 
which are responsible for the synchrotron radiation in the radio 
to submillimeter wavelengths. Our model calculations predict (Fig.~3) that the 
bremsstrahlung flux of these electrons should be observed at least at energies
between 10  MeV and 30 GeV by the future GLAST detector. It is   
also possible that instruments like INTEGRAL or Astro-E  
 will be able to detect the flux of 
bremsstrahlung photons at energies $\geq 100\,\rm keV$ (produced  
predominantly in the compact radio structures)  from Cas~A. 
In that case one expects to see a profound hardening of the radiation spectrum above
100 keV. But this effect could be detected  
only if the mean magnetic field $B_1$  would not significantly 
exceed 1\,mG, since 
otherwise the fluxes of soft $\gamma$-rays would fall below the limits of 
detectability of these instruments.

An important prediction of the spatially inhomogeneous model  
is that the spectra of accelerated electrons in Cas~A correspond to a source
function with a rather hard power-law index $\beta_{\rm acc}\sim 2.2$ (or even
slightly harder), in agreement with predictions for efficient
shock acceleration, and not a value of $\beta_{\rm acc}\sim 2.5$ as presently 
often assumed on the basis of 
 the mean spectral index of the observed radio fluxes $\alpha\approx 0.77$. 
This prediction of the model can be directly tested by future detection
of the $\gamma$-ray fluxes by GLAST in the energy region around $1\,\rm GeV$.
At these energies the total flux is dominated by
the bremsstrahlung emission of electrons which escape from the compact 
acceleration regions into the surrounding shell, resulting in
a power law distribution of electrons in the shell with an index 
$\beta_2=\beta_{\rm acc}$. Therefore, the spectral shape 
of the radiation detected at these energies will give direct information
on the spectrum of acceleration in Cas~A. The intensity of this 
radiation will give a 
rather robust estimate of the mean magnetic field $B_2$ 
 in the shell regions responsible for the diffuse `radio  plateau' emission.
This would allow also a rather good estimate of the mean magnetic field 
in the compact radio structures, $B_1\sim 4\,B_2$.  

Very important information on the magnetic field and chemical abundance 
of the gas in the compact radio structures can be derived by measuring 
the radiation fluxes at energies below 100\,MeV. Although the expected 
angular resolution of GLAST may be insufficient to
resolve the radio ring or radio knots, our study shows that most of the
flux at these energies should be produced in these compact structures. 
Therefore,
in combination with the known radio fluxes at low frequencies 
$\nu \leq 40\,\rm GHz$, 
the $\gamma$-ray spectrum of Cas~A at energies 10-100\,MeV
will provide an important information on the product of the gas parameters 
$\overline{C_{\rm Z, 1}}$ and $n_{\rm H, 1}$ in those structures. 
Adding also the information to be expected from the future  high angular
resolution measurements of the fluxes of thermal X-rays from the compact 
radio structures, even in the case of absence of line emission features it 
would
be possible to disentangle  $n_{\rm H,1}$ and $\overline{C_{\rm Z,1}}$.

A synchrotron origin of the X-ray fluxes above 10\,keV implies acceleration
of the electrons beyond 10\,TeV. If so,  we  could expect also 
noticeable fluxes of VHE $\gamma$-rays. These fluxes depend very sensitively
on the mean magnetic field $B_1$ in the compact regions of particle 
acceleration --  the bright radio ring  which is presumably connected with 
the reverse shock, and the radio knots --, and the field $B_2$ in the diffuse 
radio plateau. The model prediction for these fields (based on interpretation
of synchrotron fluxes only) corresponds to $B_1\sim (1-2)\,\rm mG$ and 
$B_2\sim B_1/4$.
For the case of low magnetic fields,
$B_1\simeq 1\,\rm mG$, the integral flux $I(> E)$ above 1 TeV 
makes about $5\,\%$ of the Crab Nebula flux, and about  $(7-8)\,\%$
above $300\,\rm GeV$. The 
$\gamma$-ray spectrum produced by electrons due to bremsstrahlung and
IC radiation is rather unusual: it is moderately hard at $E\leq 
100\,\rm  GeV$, with a photon index $\alpha_{\rm dif}\simeq 2.1-2.3$, and  
quickly steepens to $\alpha_{\gamma}> 3$ 
already at $E\simeq 1\,\rm TeV$ (see Fig.4).
An increase of the magnetic field  by only a factor of 2 results in a dramatic 
drop of the TeV $\gamma$-ray fluxes by one order of magnitude. Nevertheless, 
due to their high flux sensitivity and low detection energy threshold 
$E_{\rm th}\leq 100\,\rm GeV$ 
(Aharonian et al. 1997), the IACT arrays should be able to detect
the fluxes of VHE radiation of Cas~A even in the case  of   
high magnetic fields  $B_1\simeq 2\,\rm mG$. The preliminary detection of
Cas~A by the less sensitive IACT system of HEGRA, 
reported by P\"uhlhofer et al (1999), strengthens this expectation.

In this regard, it is significant that in Cas~A the expected 
 $\gamma$-ray fluxes of hadronic origin should readily be distinguished
from $\gamma$-rays of electronic origin due to the much steeper electronic  
spectra at energies above a few hundred GeV (see Fig.4). 
The spectral information on the VHE
$\gamma$-ray flux can be obtained by the future VERITAS array, and
might be also accessible to the HEGRA IACT system (depending on the magnetic
fields $B_1$ and $B_2$) for $\gamma$-rays of the electronic origin,
or if the total energy of relativistic protons in Cas~A is not 
less than $W_{\rm p}^{\rm (max)}\simeq 2\times 10^{49}\,\rm erg$.
In both cases the predicted flux above 500 GeV is not much less than
$(0.05-0.1)$ `Crab' (see Fig.~7). 
Note that the current upper flux limits reported by CAT (Goret et al. 1999)
and Whipple (Lessard et al. 1999) telescopes put an upper limit to the
total energy of relativistic protons in Cas~A of   
$W_{\rm p}^{\rm (max)} \leq 5\times 10^{49}\,\rm erg$ 
(assuming acceleration of protons beyond TeV energies).
 A non-detection of  
$\gamma$-ray fluxes in the region above 100\,GeV
by future IACT arrays would imply
 that the efficiency 
of hadron acceleration in Cas~A does not exceed the efficiency of 
electron acceleration.

\begin{acknowledgements} The work of AMA has been supported  
 through the Verbundforschung
Astronomie/Astrophysik of the German BMBF under the grant No. 05-2HD66A(7).
\end{acknowledgements}

\end{document}